\documentclass{article}

\usepackage[left=3.15cm,right=3.15cm]{geometry}
\usepackage{amsfonts}
\usepackage{amsmath}
\usepackage{amssymb}
\usepackage{amsthm}
\usepackage{bm}
\usepackage{mathtools}
\usepackage{mathrsfs}
\usepackage{graphicx}
\usepackage{color}
\usepackage{xcolor}
\usepackage{url}
\usepackage{hyperref}

\newtheorem{theorem}{Theorem}

\newtheorem{remark}{Remark}

\newcommand{\BibTeX}{B\kern-.05em{\sc i\kern-.025em b}\kern-.08em\TeX}
\newcommand{\argmax}{\mathop{\rm arg~max}\limits}

\usepackage[linesnumbered,ruled,vlined]{algorithm2e}
\SetKwInput{KwInput}{Input}                
\SetKwInput{KwOutput}{Output}              

\usepackage[numbers,square]{natbib}

\usepackage{authblk}
\title{Deep Mapper: Efficient Visualization of Plausible Conformational Pathways}

\author[1]{Ziyad Oulhaj\footnote{This study is based on a joint-work at Fujitsu with the mentor H.~Kurihara.}}
\author[2]{Yoshiyuki Ishii}
\author[2]{Kento Ohga}
\author[2]{Kimihiro Yamazaki}
\author[2]{Mutsuyo Wada}
\author[2]{Yuhei Umeda}
\author[2]{Takashi Katoh}
\author[2,3]{Yuichiro Wada}
\author[2]{Hiroaki Kurihara\footnote{Corresponding author. E-mail: \texttt{h\_kurihara@fujitsu.com}}}

\affil[1]{\normalsize
            Nantes Université, École Centrale Nantes, CNRS, Laboratoire de Mathématiques Jean Leray, UMR~6629, France
            }
\affil[2]{\normalsize
            Fujitsu Limited, Japan
            }
\affil[3]{\normalsize
            RIKEN Center for Advanced Intelligence Project, Japan
            }

\date{}

\begin{document}

\maketitle

\begin{abstract}
Acquiring plausible pathways on high-dimensional structural distributions is beneficial in several domains. For example, in the drug discovery field, a protein conformational pathway, i.e. a highly probable sequence of protein structural changes, is useful to analyze interactions between the protein and the ligands, helping to create new drugs. Recently, a state-of-the-art method in drug discovery was presented, which efficiently computes protein pathways using latent variables obtained from an isometric auto-encoding of the space of 3D density maps associated to protein conformations. However, our preliminary experiments show that there is room to significantly reduce the computing time. In this study, we use the Mapper algorithm, which is a Topological Data Analysis method, and present a novel variant to extract plausible conformational pathways from the isometric latent space with comparatively short running time. The extracted pathways are visualized as paths on the resulting Mapper graph. The methodological novelties are described as follows: firstly, the filter function of the Mapper algorithm is optimized so as to extract the pathways via minimization of an energy loss defined on the Mapper graph itself, while filter functions taken in the classical Mapper algorithm are fixed beforehand. The optimization is with respect to parameters of a deep neural network in the filter. Secondly, the clustering method, which defines the vertices and edges of the Mapper graph, of our algorithm is designed by incorporating domain prior knowledge to assist the extraction. In our numerical experiments, based on an isometric latent space built on the common 50S-ribosomal dataset, the resulting Mapper graph successfully includes all the well-recognized plausible pathways. Moreover, our running time is much shorter than the above state-of-the-art counterpart.
\end{abstract}

\section{Introduction}
\label{sec:intro}
Acquiring plausible pathways on a high-dimensional structural distribution is beneficial in several domains, such as food~\cite{yan2021effects} and drug discovery~\citep{Motta2022-an}. In those domains, the pathway (a.k.a. conformational pathway; see~\citep{yamazakiCryoTWIN}) is expressed by a finite sequence of chemical structures, and the structure is represented by either an all-atom model~\citep{fuchigami2011protein} or a 3D density map~\citep{yamazakiCryoTWIN}. 
The benefit in the food domain is, for example, that the plausible conformational pathways are useful to analyze interactions between antioxidants and proteins, leading to the understanding of functionality or nutritional properties of proteins~\citep{yan2021effects}.
Moreover, the benefit in the drug discovery domain is that those pathways are useful to analyze interactions between proteins and ligands, and the analysis can lead to the development of new drugs~\citep{Motta2022-an}. Because of this usefulness, several authors have proposed a method to construct such conformational pathways~\citep{fuchigami2011protein,Barroso2020Understanding,wu2022visualizing,kinman2022uncovering,yamazakiCryoTWIN}.

As a brief review of the studies above, \citet{yamazakiCryoTWIN} have proposed a protocol in the drug discovery domain for constructing a plausible protein conformational pathway as a sequence of 3D density maps
from a set of 2D projection protein images collected via cryo-Electron Microscopy (cryo-EM)~\citep{Earl2017-if}, based on their auto-encoder named cryoTWIN. 
The auto-encoder is trained by cryo-EM images; the trained auto-encoder predicts the corresponding protein 3D density map from the latent variable. CryoTWIN captures continuous structural change of the target protein via a latent distribution having a closed form. The latent space is theoretically guaranteed to be isometric to the space of 3D density maps, if the training dataset holds the manifold assumption~\citep{10.5555/1841234}. 
We note that the isometric latent distribution is efficient, since it is a low-dimensional equivalent expression to a distribution of 3D density maps. In numerical experiments, \citet{yamazakiCryoTWIN} reproduced four well-recognized 50S-ribosomal pathways by~\citet{davis2016modular} using a trained cryoTWIN by the 50S-ribosomal cryo-EM images. The reproducing process consists of the following two steps: 
(i) generate several paths via their proposed pathway computing algorithm (see~\citep[Algorithm~1]{yamazakiCryoTWIN}) using the isometric latent distribution in the trained cryoTWIN, and then aggregate the paths,
(ii) evaluate quantitatively and qualitatively whether the aggregated pathway is consistent with one of the four plausible pathways.
The plausibility of the four pathways is considered high in~\citep{yamazakiCryoTWIN}, since~\citet{davis2016modular} constructed the four pathways with heavy manual labor and standard biological tools; see visualized four pathways in~\citep[Figure~7]{davis2016modular}. \citet{yamazakiCryoTWIN} report that they could construct the plausible conformational pathways with shorter running time compared to the counterpart studies such as~\citet{kinman2022uncovering}.

In our preliminary experiments, following~\citet{yamazakiCryoTWIN}, we try to reproduce the four 50S-ribosomal pathways of~\citep{davis2016modular}, since the detailed information of the reproduction are not provided. In the experiments, we take 5 hours after training cryoTWIN by the 50S-ribosomal images. The initial 4 hours and the remaining 1 hour are from the above two steps (i) and (ii), respectively. The reason of the 4 hours in step (i) is that Algorithm~1 of~\citep{yamazakiCryoTWIN} is conducted for all pairs with significant two latent variables (number of the pairs is around 400), while the running time of Algorithm~1 for each pair is not short. We emphasize that our computational environment is the same as~\citep{yamazakiCryoTWIN}. Further details of the preliminary experiments are deferred to Section~\ref{subsec: setting}.
The preliminary experiments imply that there exists an innovation room to make step (i) more efficient in terms of running time. 

Considering the background, the goal of this study is to design a computationally efficient algorithm that returns a graph, taking as input the latent variables in an isometric latent space to the conformational space, while the graph achieves the following two conditions: (a) the vertex corresponds to a conformation and the edge expresses similarity of two conformations, and (b) a set of paths on the graph includes plausible conformational pathways.
In this study, we assume that the latent variables are obtained from an encoder of cryoTWIN taking as input cryo-EM images. Therefore, the conformation is represented by a 3D density map.

\begin{figure}[!t]
\centering
\includegraphics[width=\linewidth]{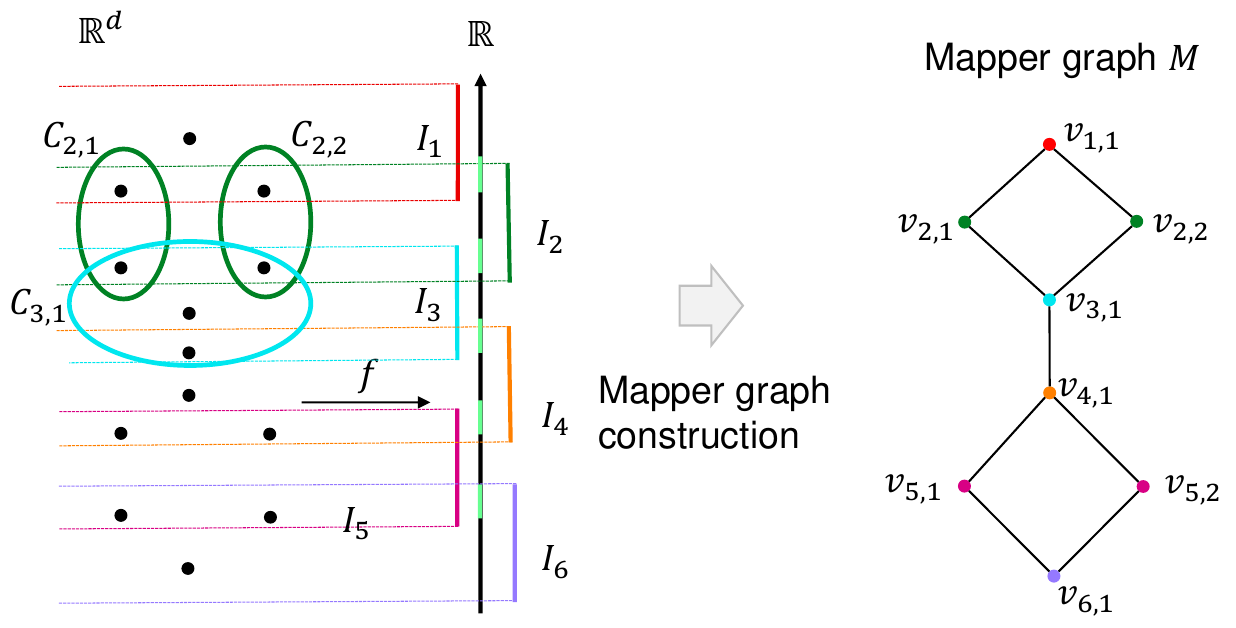}
\caption{
Example of the Mapper graph construction: Firstly, the thirteen points in the input space of $\mathbb{R}^d$ are mapped to $\mathbb{R}$ by the filter function $f$. In $\mathbb{R}$, the mapped points are covered by a set of the intervals $(I_s)_{1\leq s\leq S}, S=6$ that overlap consecutively in a certain proportion $r \in (0,1)$; see the five lime green intervals for the overlaps. Secondly, a clustering algorithm $\mathcal{A}_C$ is applied to each $f^{-1}(I_s)$, where $C$ denotes the the maximum number of clusters, and $C=2$. The two green colored groups $\mathcal{C}_{2,1}$ and $\mathcal{C}_{2,2}$ are the clustering result to $f^{-1}(I_2)$. Thirdly, based on the clustering results, the Mapper graph $M$ is constructed. In $M$, for example, the vertex $v_{2,1}$ represents the cluster $\mathcal{C}_{2,1}$, and there exists an edge between $v_{2,1}$ and $v_{3,1}$ since two clusters $\mathcal{C}_{2,1}$ and $\mathcal{C}_{3,1}$ share a point in the overlapped region.
}
\label{fig:illustration of mapper graph construction}
\end{figure}

To achieve this goal, we focus on the Mapper algorithm, which returns a visualization of topological features of high dimensional datasets in a short running time. 
In the Mapper graph, vertices and edges express a cluster in the input space and the similarity of two clusters, respectively; see Figure~\ref{fig:illustration of mapper graph construction} for how to construct the Mapper graph from a set of input data points. 
Since we employ cryoTWIN, the condition (a) is not challenging: we can identify a representative 3D density map with a vertex on the Mapper graph by applying the isometric decoder to a centroid of the cluster in the latent space. However, 
the condition (b) is challenging, because it is not trivial what clustering algorithm and filter function make the Mapper graph visualize the plausible pathways. Our other preliminary experiments show that commonly used clustering algorithms and filter functions do not work at all to achieve (b); see details in Section~\ref{subsec: setting}. 

We summarize our two main contributions in this study as follows:
\begin{enumerate}
    \item We propose a variant of the Mapper algorithm, where the clustering algorithm is designed based on the domain prior-knowledge, and the filter function is optimized using input data so as to achieve the condition (b). The filter is parameterized by a deep neural network, and the parameters are optimized via 
    minimizing a MaxFlux-objective-based energy loss on the Mapper graph, which is also parameterized by the neural network; see MaxFlux objective in~\citep{huo1997maxflux}. Since the minimizers of the original MaxFlux loss are known to be optimal reaction pathways~\citep{huo1997maxflux}, minimization of the energy loss can extract plausible conformatinal pathways from the latent space onto the Mapper graph. To the best of our knowledge, we constitute the first use of a deep neural network as a Mapper filter function, taking advantage of the universal approximation theorem~\citep[Section~20]{10.5555/2621980}. We also theoretically analyze the energy loss, when the size of the input data goes to infinity. 
    
    \item We empirically prove the efficiency of our proposed algorithm using the common 50S-ribosomal dataset. The resulting Mapper graph by our method includes all the four plausible conformational pathways by~\citet{davis2016modular}, while our running time is much shorter than the counterpart time in the step (i) of~\citep{yamazakiCryoTWIN}.
\end{enumerate}

\section{Related Work}
\label{sec:related}
We review cryoTWIN~\citep{yamazakiCryoTWIN} in Section~\ref{subsec: Review of CryoTWIN}, since we employ it as preprocessing technique to obtain latent variables in our numerical experiments. 
In Section~\ref{subsec: review mapper alg}, 
we first introduce the Mapper algorithm, and then review existing uses of the Mapper algorithm designed for biological applications.

\subsection{CryoTWIN}
\label{subsec: Review of CryoTWIN}
\citet{yamazakiCryoTWIN} proposed a method to compute a plausible protein conformational pathway from the single particle cryo-EM images.
In their method, first, an isometric latent space to a space of 3D density maps is built via training an auto-encoder named cryoTWIN. Then, a plausible protein conformational pathway is computed as a sequence of 3D protein density maps, utilizing the isometric latent space.

CryoTWIN consists of an encoder $h_{\zeta}$, a decoder $g_{\xi}$, and a latent distributional model $P_\psi$, where $\zeta, \xi, \psi$ are trainable parameters. The encoder outputs the latent variable $z$ as input of a Fourier transformed cryo-EM image $x$. The latent model is defined as a Gaussian Mixture Model (GMM) $P_\psi, \psi = \{(\pi_k, \mu_k, \Sigma_k)\}_{k=1}^{K}$, where $\pi_k, \mu_k,$ and $\Sigma_k$ represent $k$-th Gaussian's weight, mean, and variance, respectively, and $\sum_{k=1}^K w_k =1\;\forall k;w_k\geq0$. 
Let $x_i$ be a Fourier transformation of $i$-th cryo-EM image.
The training objective to obtain the optimized parameters $\zeta^\ast,\xi^\ast,\psi^\ast$ is as follows:
\begin{equation}
\label{eq: cryotwin objective}
    \arg\min_{\zeta, \xi, \psi}\frac{1}{N}\sum_{i=1}^N
    \mathbb{E}_{\varepsilon}\biggl\lbrack
    \left\|W\odot \left(x_i - \hat{x}_{{z}_{i} + \varepsilon}\right)\right\|_{2}^{2} - \beta\log P_\psi ({z}_{i})
    \biggr\rbrack,
\end{equation}
where 
the symbols $N, \varepsilon, \odot$, and $\beta$ are the number of cryo-EM images, random noise, Hadamard product, and the positive hyper-parameter, respectively. In addition, $\hat{x}_{{z}_{i} + \varepsilon} = g_{\xi} (z_i + \varepsilon, \hat{R}_i)$, $z_i = h_{\zeta}(x_i)$, $\hat{R}_i$ is the corresponding estimated pose orientation to $x_i$, and $\hat{x}_{{z}_{i} + \varepsilon}$ is the predicted $2$D Fourier image in the $3$D Fourier volume. The symbol $W$ is a weight matrix for introducing the isometricity; see~\citep[Appendix~A]{yamazakiCryoTWIN}. CryoTWIN is inspired by another auto-encoder namely RaDOGAGA~\citep{DBLP:conf/icml/KatoZSN20}, which also builds an isometric latent space to the original space. 
After the training, 
for a latent variable $z$, 
the corresponding 3D density map $\hat{V}_z$ is reconstructed by the trained decoder $g_{\xi^\ast}$.

The isometricity to a space of $3$D density maps enables us to compute 
plausible conformational pathways via the trained GMM $P_{\psi^\ast}$ using the decoder $g_{\xi^\ast}$. In~\citep{yamazakiCryoTWIN}, the pathway computation algorithm is proposed; see also Algorithm~1 of~\citep{yamazakiCryoTWIN}. This algorithm requires two means $\mu_i^\ast$ and $\mu_j^\ast$ as start and end points of the pathway. Then, a MaxFlux path~\citep{huo1997maxflux} on $P_{\psi^\ast}$ between $\mu_i^\ast$ and $\mu_j^\ast$ is approximately computed using greedy optimization technique. The following minimization problem is used in the greedy optimization at $t+1$-th iteration to compute the $t+1$-th latent variable ${z}^{i\to j}(t+1)$:
\begin{equation}
\label{eq: greedy pathway computation}
    {z}^{i\to j}(t+1) = \underset{ \{\textrm{z}\} }{\operatorname{arg\;min}} \frac{\|\textrm{z} - {z}^{i \to j}(t) \|_2}{P_{\psi^\ast}(\textrm{z})},   
\end{equation}
where $\{\textrm{z}\}$ is a set of the candidate latent variables. The path on $P_{\psi^\ast}$ is expressed as a sequence of the latent variables: $(\mu_i^\ast,\dots ,z^{i \to j}(t),\dots ,\mu_j^\ast)$. Thereafter, 
by decoding each element in the sequence $(\mu_i^\ast,\dots ,z^{i \to j}(t),\dots ,\mu_j^\ast)$ using $g_{\xi^\ast}$, the algorithm outputs  $(\hat{V}_{\mu_i^\ast},\dots ,\hat{V}_{z^{i \to j}(t)},\dots ,\hat{V}_{\mu_j^\ast})$: a sequence of $3$D density maps.

\subsection{Mapper Algorithm}
\label{subsec: review mapper alg}
\paragraph{Definition of the Mapper algorithm:} Let $X$ be a topological space and let $f\colon X \to \mathbb{R}$ be a continuous function called a \textit{filter function}. We define an equivalence relation between two elements $x$ and $y$ in $X$ by $x\sim_{f} y$ if and only if $x$ and $y$ are in the same connected component of $f^{-1}(a)$ for some $a$ in $f(X)$. Then the \emph{Reeb graph} 
$\mathcal{R}_f(X)$ of a topological space $X$ computed with a filter function $f$ is defined to be the quotient space $X/\sim_{f}$. The \emph{Mapper graph}, introduced in \cite{singh2007topological}, is a statistical version of the Reeb graph consisting of a computable approximation. It is discrete in the sense that it is computed on a finite metric space $(\mathbb{X}_n=\{x_0,\dots,x_n\} \subseteq X,d_X)$ with a continuous filter function $f$, where $d_X$ is a metric on $X$. 

The procedure to compute a Mapper graph $M$ is the following three steps~\citep{introduction_tda} 
\begin{description}
    \item[\textit{Step~1:}] Cover the range of values $f(\mathbb{X}_n)$ with a set of intervals $(I_s)_{1\leq s\leq S}$ that overlap consecutively in a certain proportion $r \in (0,1)$. 
    \item[\textit{Step~2:}] Apply a clustering algorithm $\mathcal{A}_C$ to each pre-image $f^{-1}(I_s)$, $s\in\{1,\dots,S\}$, where $C$ is the maximum number of clusters. This produces the \textit{pullback cover} $\mathscr{C}$ of $\mathbb{X}_n$, where 
    $\mathscr{C} = \{ \mathcal{C}_{1, 1}, \dots , \mathcal{C}_{1, C_{1}}, \dots , \mathcal{C}_{s, 1}, \dots ,  \mathcal{C}_{s, C_{s}}, \dots, \mathcal{C}_{S, 1}, \dots , \mathcal{C}_{S, C_{S}} \}$ and $C_s \leq C$ for any $s$, and $\mathcal{C}_{s,c}$ denotes the $c$-th cluster of $f^{-1}(I_s)$.
    
    \item[\textit{Step~3:}] The Mapper graph $M$ is the $1$-skeleton of the nerve complex of 
    $\mathscr{C}$.
    It is a graph with a vertex $v_{s,c}$ for each $\mathcal{C}_{s,c}$, and an edge between two vertices $v_{s,c}$ and $v_{s',c'}$ if and only if $\mathcal{C}_{s,c} \cap \, \mathcal{C}_{s',c'}\neq \emptyset$. 
\end{description}
See Figure~\ref{fig:illustration of mapper graph construction} for illustration of the three steps. 
In Appendix~\ref{sec: complements for mapper graph}, we provide complementary information.

\begin{algorithm}[!t]
\caption{Deep Mapper graph}
\label{alg:generic}
    \KwInput{
    \begin{enumerate}
        \item $\{z_i\}_{i=1}^N$: Set of latent variables in $\mathbb{R}^d$ from an isometric auto-encoder to the structural space,
        \item $P_{\psi}(z)$: The latent distribution model by GMM, where $\psi$ is the GMM parameters,
        \item $\Bar{K}$: Number of significant Gaussian components,
        \item $\kappa$: Number of neighbors,
        \item $S\;\textrm{and}\;r \in (0,1)$: Number of intervals and the overlap rate,
        \item $C$: Maximum number of clusters,
        \item $\lambda$: Positive fixed value,
        \item $g$: Isometric decoder from a latent variable to the structure.
    \end{enumerate}}
    \KwOutput{Optimized Mapper graph with structures.} 
    Using $\kappa$, build a kNN graph $G=(\mathcal{V}, \mathcal{E})$ on $\{z_i\}_{i=1}^N$, where $\mathcal{V}=\{z_i\}_{i=1}^N$ and $\mathcal{E}$ is the set of the edges.

    Select the significant Gaussian components in $P_{\psi}(z)$. Then, using the significant components and $C$, define a clustering algorithm $\mathcal{A}_C$ to each pre-image $f^{-1}_\theta (I_s)$, where $I_s$ is the $s$-th ($1\leq s \leq S$) interval in $\mathbb{R}$; see the interval in \textit{Step~1} of Section~\ref{subsec: review mapper alg}. 
    The clustering algorithm is designed to provide a representative variable to each cluster. 
    Also, based on the selected components, define the filter function $f_{\theta}:\mathbb{R}^d \to \mathbb{R}$, which is parameterized by trainable parameters $\theta$ in a deep neural network. 
    See further details in Section~\ref{subsubsec: def of clustering alg.}.

\For{each epoch}
    {
    Following \textit{Step~1} to \textit{Step~3} in Section~\ref{subsec: review mapper alg}, construct a Mapper graph $M_\theta$ on $\{z_i\}_{i=1}^N$ by using $S$, $r$, $f_\theta$, and $\mathcal{A}_{C}$.
    
    Compute the energy loss $L_\theta$ on $M_\theta$ using $P_\psi$; see Section~\ref{subsubsec: def of the energy loss}.
    
    Compute a regularization term $\mathrm{Reg}_{\theta}(G)\coloneqq \lambda \sum_{(i, j) \in \mathcal{E}} (f_{\theta}(z_i)-f_{\theta}(z_j))^2$ on $G$.
    
    $L_\theta \leftarrow L_\theta + \mathrm{Reg}_{\theta}(G)$, and then minimize $L_\theta$ with respect to the parameters $\theta$.
    }

    For each representative latent variable in the optimized Mapper graph, compute the corresponding structure using the decoder $g$. See Section~\ref{subsubsec: def of clustering alg.} for the representative latent variables.
    
\Return $M_\theta$ with the structures.
\end{algorithm}

\paragraph{Review of Mapper algorithms in biology applications:} 
The Mapper graph can capture various topological features in data which has a complicated structure. 
It is also known that the Mapper algorithm is more robust to the distance compared to non-linear dimension reduction or geometric embedding methods; see~\citep{yao2009topological}.
As examples of applications of the Mapper, \citet{yao2009topological} applied the Mapper algorithm to characterize transient intermediates or transition states which are quite crucial for the description of biomolecular folding pathways. In \citep{nicolau2011topology}, a Disease-Specific Genomic Analysis is first performed on a breast cancer dataset to produce measures of deviation between tumor and normal tissue. This information is then introduced to the Mapper algorithm as a filter function. Due to its ability to conserve topological information in the dataset, the resulting graph reveals a region that corresponds to a unique mutational profile, that is otherwise scattered across different clusters in a regular clustering analysis.

In the context of single-cell RNA sequencing data analysis, \citet{wang2018topological} proposed a Mapper algorithm, whose filter function is designed via gene co-expression network analysis. As a result, the Mapper graph not only preserved the continuous nature in gene expression profiles, but also successfully separated different cell types. In \citep{imoto2023v}, the authors adopted the Mapper algorithm to data with velocity, and associated flow on edges of the resulting Mapper graph. They applied their method to single-cell gene expression and combined their method with the Hodge decomposition on a graph to enhance the interpretation of the flow on the Mapper graph. 

A relaxed and more general version of the Mapper graph, that enjoys improved stratification properties, is introduced in \cite{oulhaj2024differentiable}. It is then used to optimize parameterized filter functions for regular Mapper graphs, with respect to a topological risk based on persistent homology. This is shown to produce quality Mapper graphs for 3-dimensional shapes and single cell RNA-sequencing data.


\section{Proposed Algorithm}
\label{sec:proposed method}
Our proposed algorithm is shown in Algorithm~\ref{alg:generic}. 
Assuming that the original dataset holds the manifold assumption in a high-dimensional structural space, this algorithm requires a set of latent variables $\{z_i\}_{i=1}^N$, where $z_i$ is the encoding of the $i$-th original data point with an isometric auto-encoder. Thanks to this isometric auto-encoder such as cryoTWIN~\citep{yamazakiCryoTWIN}, we can use the low-dimensional latent space, which is equivalent to the high-dimensional structural space. Using $\{z_i\}_{i=1}^N$, our algorithm extracts plausible conformational pathways from the latent space, and visualizes them on the resulting Mapper graph. 
The extraction is achieved by an optimized filter function $f_\theta$ and our designed clustering algorithm $\mathcal{A}_C$. The optimization is via minimization of a MaxFlux objective~\citep{huo1997maxflux} inspired energy loss $L_\theta$ defined on the parameterized Mapper graph $M_\theta$. 
\begin{remark}
    As mentioned at the fourth paragraph of Section~\ref{sec:intro}, in this study, we assume cryoTWIN~\citep{yamazakiCryoTWIN} to prepare the first and ninth inputs of Algorithm~\ref{alg:generic}. However, any isometric auto-encoders to the original structural space are applicable as long as they enable us to prepare those inputs from a set of original data. Furthermore, once the latent variables are obtained from the isometric auto-encoder, we can fit a Gaussian mixture model to those variables by~\citep[Section 24.4.2]{10.5555/2621980}, and then we can prepare the second input.
\end{remark}

The definitions of our clustering algorithm $\mathcal{A}_C$ and the filter function $f_\theta$ are given in Section~\ref{subsubsec: def of clustering alg.}, while that of the energy loss is described in Section~\ref{subsubsec: def of the energy loss}.  Thereafter, we theoretically analyze Algorithm~\ref{alg:generic} in Section~\ref{subsec: theory for specific algorithm}.

\subsection{The Clustering Algorithm and Filter Function}
\label{subsubsec: def of clustering alg.}
As the second input to Algorithm~\ref{alg:generic}, let us consider a trained GMM $P_\psi (z) = \sum_{k=1}^K \pi_k \mathcal{N}(z; \mu_k, \Sigma_k)$ obtained from cryoTWIN (see Section~\ref{subsec: Review of CryoTWIN}), where $\mathcal{N}$ expresses a Gaussian distribution. 
Inspired by \textsf{Expt2} in~\citep{yamazakiCryoTWIN}, we select the significant Gaussian components: we find top $\Bar{K}$ largest weight values $\pi_k, k=1, \dots, K$, and define the corresponding Gaussian indexes as $k_j\;(j=1,\dots,\Bar{K})$, where $\pi_{k_j} \leq \pi_{k_{j+1}}$. 

Let $\mathcal{K}=\{k_j|j=1,\dots,\Bar{K}\,\&\,\forall j;\pi_{k_j} \leq \pi_{k_{j+1}}\}$. Then, 
our clustering algorithm $\mathcal{A}_C$ to pre-image $f^{-1}_{\theta}(I_s)$ consists of the following three steps. Firstly, for each $k \in \mathcal{K}$, compute $Q_k = \max_{z \in f^{-1}_{\theta}(I_s)} \pi_k \mathcal{N}(z; \mu_k, \Sigma_k)$. Secondly, define "Candidates" as a set of indexes with the top $C$ largest values in $\{Q_k\}_{k=1}^{\Bar{K}}$. Thirdly, for each $z \in f_\theta^{-1}(I_s)$, estimate the cluster label of $z$ by $\argmax_{k\in\text{Candidates}} \pi_k \mathcal{N}(z; \mu_k, \Sigma_k)$. Based on the labels, group the variables in $f_\theta^{-1}(I_s)$ as clusters. In addition, define a representative latent variable to each cluster by the mean vector of $\mu_k$, where $k \in \mathcal{K}$ denotes the estimated cluster label.

In the first step, we compute $Q_k$ for $k$-th Gaussian component to measure how much the component is related to the pre-image $f_\theta^{-1}(I_s)$ using the density of the joint distribution $\pi_k \mathcal{N}(z; \mu_k, \Sigma_k)$. In the second step, we pick the most related $C$ components to define the set "Candidates". In the third step, we use Bayes' theorem (see~\citep[Section~24.5]{10.5555/2621980}) to estimate cluster label of $z$ from "Candidates". Note that we can compute an associated structure to each cluster via decoding $\mu_k$ by the eighth input $g$ in Algorithm~\ref{alg:generic}; see the decoding in Section~\ref{subsec: Review of CryoTWIN}.
Additionally, we use the prior knowledge to design the clustering: a set of conformations obtained by decoding mean vectors of significant Gaussian components in cryoTWIN is empirically almost equivalent to a set of important conformations.

Next, the filter function is given by a neural network that acts on a feature transform of the latent space. Specifically, the filter function has the form of $f_{\theta}(z)=DNN_\theta(T(z))$. The symbol $DNN_\theta$ is a deep neural network with trainable parameters $\theta$. In addition, $T(z)$ is a vector via the map $T:\mathbb{R}^d \to \mathbb{R}^{|\Bar{K}|}$, and for $j=1,\dots ,\Bar{K}$, the $j$-th element of $T(z)$ 
is defined by $(z-\mu_{k_j})^\top\Sigma^{-1}_{k_j}(z-\mu_{k_j}), k_j \in \mathcal{K}$, i.e., the map $T$ is a characterization of $z$ by the top $\Bar{K}$ significant Gaussian components with Mahalanobis' distance~\citep{mahalanobis1936generalized}.

\subsection{The Energy Loss and Minimization}
\label{subsubsec: def of the energy loss}
Let $M_\theta = (\Tilde{\mathcal{V}}_\theta, \Tilde{\mathcal{E}}_\theta)$ denote the resulting Mapper graph after line 4 of Algorithm~\ref{alg:generic}. 
Here, $\Tilde{\mathcal{V}}_\theta$ is a set of vertexes with the Mapper graph, and $\Tilde{\mathcal{E}}_\theta$ is a set of the edges. Since the vertex set can be identified with a set of clusters $\{\mathcal{C}\}$ defined by $\mathcal{A}_C$, we express the edge by $(\mathcal{C}, \mathcal{C}^\prime)\in \Tilde{\mathcal{E}}_{\theta}$.

For computing the energy loss $L_\theta$ on the Mapper graph $M_\theta$, set $L_\theta$ and $B$ to zero and a natural number, respectively. Then, for each edge $(\mathcal{C}, \mathcal{C}^\prime)\in \Tilde{\mathcal{E}}_{\theta}$, 
compute the edge-wise energy loss $\ell_\theta (\mathcal{C}, \mathcal{C}^\prime)$ by the following three steps. Firstly, sort the latent variables in $\mathcal{C}\cup \mathcal{C}^\prime$ based on the filter value in ascending order. Secondly, for $b=1,\dots ,B$,  let $\iota(b) \in \{1,\dots ,N\}$ denote an index of the $\lfloor\frac{b (\mid \mathcal{C}\cup \mathcal{C}^\prime \mid-1)}{B}\rfloor$-th latent variable in the order. Thirdly, the edge-wise energy loss is given by 
$\sum_{b=1}^{B-1} \text{LDA}(z_{\iota(b)}, z_{\iota(b+1)},f_{\theta},\nabla f_{\theta}(z_{\iota(b)})) / P_{\psi}\left(\frac{z_{\iota(b)}+z_{\iota(b+1)}}{2}\right).$ At last, the energy loss $L_\theta$ is defined by $L_\theta = \sum_{(\mathcal{C}, \mathcal{C}^\prime)\in \Tilde{\mathcal{E}}_{\theta}} \ell_\theta (\mathcal{C}, \mathcal{C}^\prime)$. 

The LDA is an abbreviation of \textit{Local Distance Approximation}. As the name suggests, the value of $\mathrm{LDA}(z, z', f, \nabla f (z))$ is an approximation to the distance $\|z' - z\|_2$. The motivation to introduce LDA is to make $\ell_\theta (\mathcal{C}, \mathcal{C}^\prime)$ differentiable w.r.t. $\theta$. Let us assume $z \approx z'$, and let $z_i$ denote the $i$-th element in $z$. Additionally, we define a $d$-dimensional vector of $z_{-i}$ by replacing the $i$-th element in $z$ with $z'_i$. By the definitions, $\text{LDA}(z, z', f, \nabla f (z))$ is given by
$\left\|\left( \frac{f(z_{-1}) - f(z)}{\nabla f(z)_1},\dots,\frac{f(z_{-i}) - f(z)}{\nabla f(z)_i},\dots,\frac{f(z_{-d}) - f(z)}{\nabla f(z)_d} \right)\right\|_2$, where $\nabla f(z)_i$ is the $i$-the element in the gradient of $f$ at $z$; see details of LDA in Appendix~\ref{sec: details of lda}.

From the definition of LDA, the energy loss $L_\theta$ is approximately equal to $\sum_{(\mathcal{C}, \mathcal{C}^\prime)\in \Tilde{\mathcal{E}}_{\theta}} \sum_{b=1}^{B-1} \|z_{\iota(b+1)} - z_{\iota(b)}\|_2 / P_{\psi}\left(\frac{z_{\iota(b)}+z_{\iota(b+1)}}{2}\right)$. Thus, minimization of $L_\theta$ w.r.t. $\theta$ can be considered as the MaxFlux objective on the Mapper graph $M_\theta$; see~\citep[Equation~(17)]{huo1997maxflux} for the original objective. Since the minimizers of the original MaxFlux objective are known to be optimal chemical reaction pathways~\citep{huo1997maxflux}, i.e., a kind of plausible conformational pathways on a chemical structural distribution, the minimization of $L_\theta$ has a potential to extract the plausible pathways from the latent distribution $P_\psi$, leading to  visualization of the pathways on $M_\theta$.

Note that only the LDA related parts are parameterized by $\theta$ in $\ell_\theta (\mathcal{C}, \mathcal{C}^\prime)$, whereas $P_\psi$ related parts are not. Indeed, it is natural to also parameterize the latter parts by $\theta$ since they are components in the original MaxFlux objective. However, since the way of the parameterization is not trivial, we only parameterize the former parts in this study.

As line 7 of Algorithm~\ref{alg:generic} shows, we add a regularization term $\textrm{Reg}_{\theta}(G)$  to $L_\theta$ in practice, where the regularizer is computed on a k-Nearest Neighbor (kNN) graph of $\{z_i\}_{i=1}^N$; see~\citep[Section 19.1]{10.5555/2621980} for kNN graph.
In this study, we compute the exact kNN graph with brute-force search, whose time and memory complexities are $O(dN^2)$ and $O(\kappa N)$, respectively. To reduce the time complexity, the approximation method of~\citep{10.5555/3120137.3120188}, whose time complexity is $O(dN\log N)$, is available. The minimization of $\textrm{Reg}_{\theta}(G)$ helps the filter function to preserve the data manifolds of $\{z_i\}_{i=1}^N$ onto the filter space. The final loss in the line 7 is minimized using Adam optimizer~\citep{kingma2014adam} with He initialization~\citep{7410480}.

\subsection{Theoretical Analysis}
\label{subsec: theory for specific algorithm}
We study the properties of the energy loss $L_\theta$, when $B$ and $N$ of Section~\ref{subsubsec: def of the energy loss} go to infinity. In the following paragraphs, we prove two properties below: firstly, the discretized edge-wise energy loss $\ell_\theta (\mathcal{C}, \mathcal{C}')$ converges to the continuous counterpart when $B \to \infty$ (see Equation~\eqref{eq1} in Theorem~\ref{thm}), and secondly, the discretized energy loss of $L_\theta$ converges to the continuous MaxFlux loss in the Mapper graph when $B, N \to \infty$; see Equation~\eqref{eq2} in Theorem~\ref{thm}. For the original continuous MaxFlux loss, see~\citep[Equation~(6)]{huo1997maxflux}.

Consider the case that we are interested in an optimal conformational path between two protein conformation representations $z_0$ and $z_\infty$, where $z_0,z_\infty\in \mathbb{R}^d$ ($d\in\mathbb{N}$). 
We also assume that we know a closed form $P_{\psi} \colon \mathbb{R}^d \to [0,+\infty)$ for the probability density of the protein conformation representations, and $P_{\psi}(\mathbb{R}^d)$ is connected.

We denote by $\mathrm{PC}^0\left([0,1],\mathbb{R}^d \right)$ the set of piecewise continuous functions $f\colon [0,1]\rightarrow \mathbb{R}^d$ that satisfy $f(0)=z_0$ and $f(1)=z_\infty$. 
A map $f\in \mathrm{PC}^0\left([0,1],\mathbb{R}^d \right)$ will be called a \textit{conformational path} between $z_0$ and $z_\infty$.
For $f\in \mathrm{PC}^0\left([0,1],\mathbb{R}^d\right)$, we define its energy loss $I_f$, which is defined as the Riemann integral, as 
$I_f = \int_{0}^{1} 1/P_{\psi}(f(t)) \mathrm{d}f(t) \coloneqq \lim_{n \to \infty} \sum_{i=0}^{n-1}\Vert f(\frac{i+1}{n}) - f(\frac{i}{n})) \Vert_2 / P_{\psi}((f(\frac{i}{n}) + f(\frac{i+1}{n})) / 2)$.
The optimal energy loss for all possible conformational paths is defined as
$S^\ast \coloneqq \inf_{f\in \mathrm{PC}^0\left([0,1],\mathbb{R}^d\right)}I_f.$
We here remark the following two: first, the notation $f$ used in this section is different from a filter function $f$ appeared in Section~\ref{subsubsec: def of clustering alg.}, and second, $I_f$ can be interpreted as the energy loss on the Mapper graph if a map $f$ defines a path on the graph.

In our work, we wish to approach the optimal energy loss by using a finite $N$ samples $\{z_j\}_{j=1}^N$ of conformations, that independently follow a distribution of density $P_{\psi}$. We then discretize the energy loss using $l$ steps; the symbol $l$ corresponds to $B$ in Section~\ref{subsubsec: def of the energy loss}. Precisely, we define the \emph{stochastic loss} with $l $ steps associated to $\{z_j\}_{j=1}^N$ as 
$S_{l}^{(N)}\coloneqq \min_{\{z_i\}_{i=1}^{l-1}\subseteq\{z_j\}_{j=1}^N}\sum_{i=0}^{l-1}\left\Vert z_{i+1}-z_i \right\Vert_2 / P_{\psi}\left(\frac{z_i+z_{i+1}}{2}\right),$
with the convention $z_{l}=z_{\infty}$, where 
$\{z_j\}_{j=1}^N\subseteq\mathbb{R}^d$ is a random sample drawn independently from a distribution of density $P_{\psi}$ and $l \in\mathbb{N}$ such that $l >2$. 

The theoretical part of our work aims at proving that $S_{l}^{(N)}$ gets arbitrarily close to $S^\ast$ when $N$ and $l $ go to infinity. In order to properly define this property in terms of real sequence convergence and random variable sequence convergence, we define the \emph{discrete optimal energy loss} with $l $ steps as 
$S_{l} \coloneqq \inf_{\{z_i\}_{i=1}^{l-1}\subseteq\mathbb{R}^d}\sum_{i=0}^{l-1}\left\Vert z_{i+1}-z_i \right\Vert_2 / P_{\psi}\left(\frac{z_i+z_{i+1}}{2}\right),$
with the same convention $z_{l}=z_{\infty}$, where $l \in \mathbb{N}$ such that $l >2$.
We then have the following result regarding the different losses that we defined above.

\begin{theorem}\label{thm}
We have 
\begin{equation}\label{eq1}
S_{l}\xrightarrow[l \to \infty]{}S^\ast.
\end{equation}

Moreover, for every $l \in\mathbb{N}$ such that $l >2$, assuming that 
    (i) the density $P$ is continuously differentiable, 
    (ii) the density $P$ is strictly positive almost everywhere, and 
    (iii) we now only consider discrete paths $\{z_i\}_{i=1}^{l-1}\subseteq\mathbb{R}^d$ such that  $\forall i\in\{0,\dots,l-1\}$ $z_{i+1}\neq z_i$.
Then we have
\begin{equation}\label{eq2}
S_{l}^{(N)}\xrightarrow[N \to \infty]{a.s}S_{l}.
\end{equation}
\end{theorem}
The proof of this theorem is given in Appendix~\ref{append: proof eq1} and~\ref{append: proof eq2}.

\begin{figure*}[!t]
      \begin{minipage}[!t]{.465\linewidth}
      \centering
        \includegraphics[width=0.865\hsize]{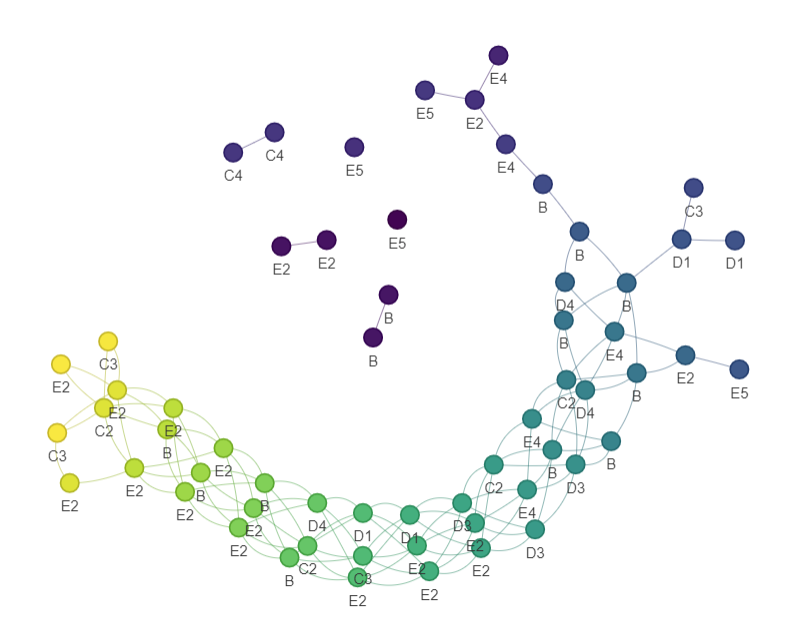}%
      \end{minipage} 
      \begin{minipage}[!t]{.55\linewidth}
      \centering
        \includegraphics[width=0.95\hsize]{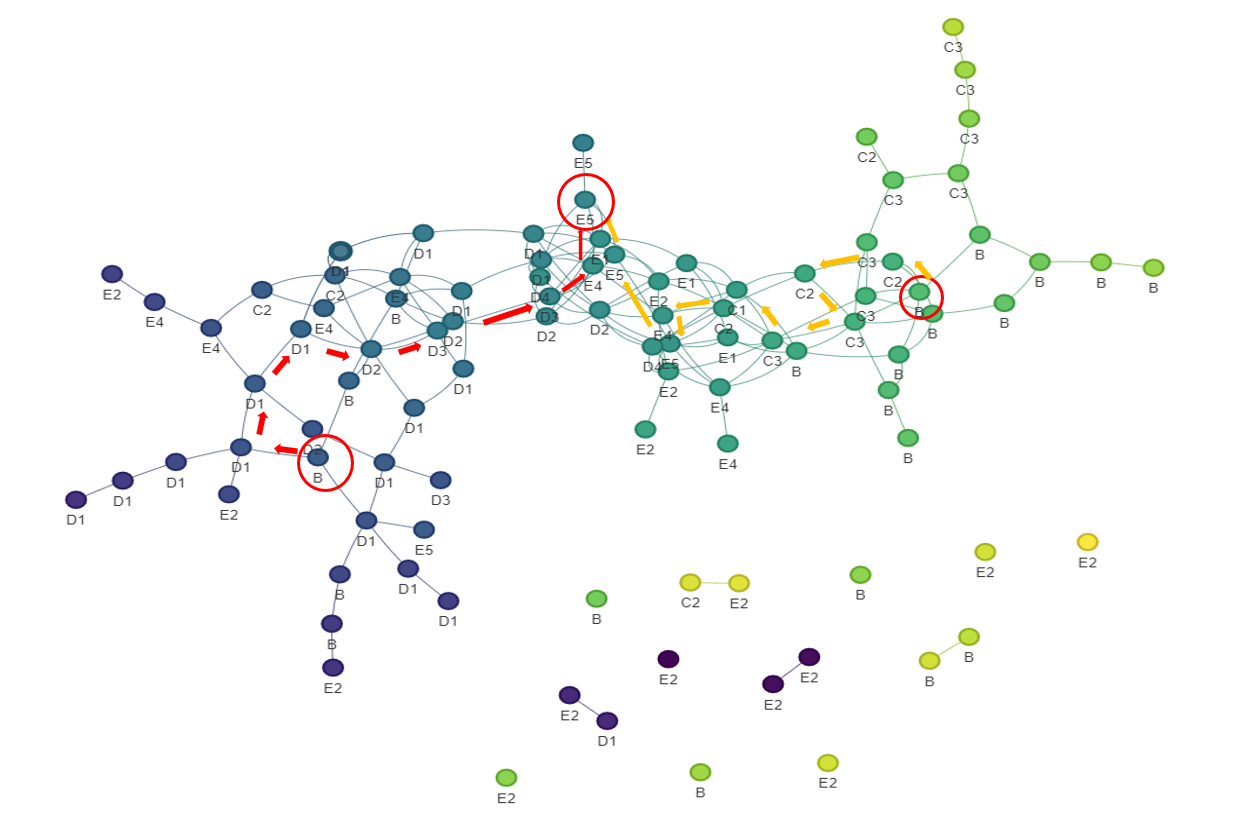}
      \end{minipage} 
\caption{
Left and right graphs: Resulting Mapper graphs of 2nd preliminary and main experiments, respectively.
In both graphs, each color of nodes corresponds to the average value over the filtered values in each cluster. 
The labels such as B on the graphs are the 50S-ribosomal structural label defined by~\citet{davis2016modular}. The red and orange arrows in the right graph are two of four well-recognized conformatinal pathways defined also in~\citep{davis2016modular}.
}
\label{fig: classical vs ours}   
\end{figure*}

\section{Numerical Experiments}
\label{sec:numerical experiments}
In this section, we introduce our numerical experiments. In Section~\ref{subsec: setting}, we describe our setting and result. Then, we discuss the result in Section~\ref{subsec: result and discussion}. Thereafter, we describe both ablation and robustness studies against our method in Section~\ref{subsec: ablation} and~\ref{subsec: robustness}, respectively.

\subsection{Setting and Result}
\label{subsec: setting}
Through all experiments, we use a set of the latent variables $\{z_i\}_{i=1}^N$, which are the encoding results of cryoTWIN~\citep{yamazakiCryoTWIN} as input of 50S-ribosomal cryo-EM images from EMPIAR-10076; see~\citep{10.1093/nar/gkac1062} for what EMPIAR is. 

In the following, we first describe the preprocessing for our experiments. Then, we introduce 1st and 2nd preliminary experiments, which motivate us to propose the deep Mapper shown in Algorithm~\ref{alg:generic}. Finally, we explain our main experiment.

\paragraph{Preprocessing:}
We download the 50S-ribosomal cryo-EM images with their estimated pose orientations from GitHub URL of~\citep{zhonge2019ribosomal}. The image size is $128\times 128$. 
We train cryoTWIN by the downloaded dataset with the training objective in Equation~\eqref{eq: cryotwin objective}. For the training, we employ the same hyperparameter values and computational resource (i.e., four NVIDIA V100 GPU accelerators with two Intel Xeon Gold 6148 processors) as~\citet{yamazakiCryoTWIN}. 
After the training, we prepare the latent variable set $\{z_i\}_{i=1}^N$ by the encoder, where $N=131899$ and $\forall i; z_i \in \mathbb{R}^8$. Then, following~\citep[\textsf{Expt2}]{yamazakiCryoTWIN}, we evaluate whether we can observe the structural labels defined by~\citet{davis2016modular} (e.g., B, C2, D1, E5,... etc) in the corresponding 3D density maps of the mean vectors of significant Gaussian components. 
The evaluation is based on PyMOL~\citep{PyMOL} and Fourier Shell Correlation (FSC) metric; see FSC in~\citep[3rd footnote]{Zhong2020Reconstructing}. In this preprocessing, we observe almost all important structural labels; see details of the quantitative evaluations by FSC in 
Appendix~\ref{subsec: FSC Evaluations in Preprocessing}.

\paragraph{1st preliminary experiment:}
The aim of this experiment is to 
measure the running time in reproducing the four pathways of~\citep{davis2016modular}, based on the above trained cryoTWIN via the protocol shown in~\citep[\textsf{Expt2}]{yamazakiCryoTWIN}. To do so, we only focus on the
mean vectors of top 30 significant Gaussian components out of 100, since the decodings of the remaining 70 mean vectors are not necessarily consistent with
well-recognized structures by~\citet{davis2016modular}.
Note that we measure the significance based on the Gaussian weight $\pi$, as~\citet{yamazakiCryoTWIN} do.
Then, (i) we generate the ribosomal path by applying the pathway computing algorithm of Algorithm~1 in~\citep{yamazakiCryoTWIN} to a pair of the significant mean vectors; see Equation~\eqref{eq: greedy pathway computation} for how to generate the path. For Algorithm~1 of~\citep{yamazakiCryoTWIN}, we use exactly the same hyperparameter values as~\citep{yamazakiCryoTWIN}. Since the number of pairs is 435, we conduct parallel computing for the pairs using the same computers in the preprocessing. At last, (ii) we aggregate the paths, and evaluate whether the aggregated path is consistent with one of the four pathways. The running time in step (i) is around 240 minutes, while the time in step (ii) is around 60 minutes.

\paragraph{2nd preliminary experiment:}
The aim of this experiment is to qualitatively evaluate whether the resulting Mapper graph, which is obtained via applying a classical Mapper algorithm to the set of latent variables $\{z_i\}_{i=1}^N$ from the preprocessing, can contain the four well-recognized 50S-ribosomal conformational pathways of~\citet[Figure~7]{davis2016modular}.
Following Steps 1 to 3 in Section~\ref{subsec: review mapper alg}, 
we employ $f_{\rm mean}$ and k-means~\citep[Section~22.2]{10.5555/2621980} for the filter function and the clustering algorithm in the classical algorithm, respectively. Here, $f_{\rm mean}$ returns the average value over $d$-elements in $d$-dimensional vector, and the number of clusters in k-means is $3$. Additionally, the numbers of intervals $S$ and overlap rate $r$ are $S=20$ and $r=0.25$, respectively. 
Moreover, we introduce a representative latent variable for each cluster as follows. Let $\{\mu_{k_j}\}_{j=1}^{25}$ denote a set of mean vectors related to the top $25$-significant Gaussian components, and let $q$ denote the cluster centroid given by k-means. 
The definition of the significance is the same as in the 1st preliminary experiment.
Then, the representative latent variable is defined by $\arg\min_{\{\mu_{k_j}\}_{j=1}^{25}}\|\mu_{k_j} - q\|_2$. Note that the structural labels corresponding to the top 25 Gaussian mean vectors do not correspond to junk structures. The resulting Mapper graph with the labels is shown in the left-hand side of Figure~\ref{fig: classical vs ours}. The running time is less than one minute using Apple M$1$ $16$ GB $8$ cores.

\paragraph{Main experiment:}
We apply our deep Mapper of Algorithm~\ref{alg:generic} to the set of latent variable $\{z_i\}_{i=1}^N$ using the trained GMM $P_\psi$ from the preprocessing. 
The aim of this experiment is not only to evaluate whether the resulting Mapper graph can contain the four plausible pathways, but also to measure the running time. We set $(\Bar{K}, \kappa, S, r, C, \lambda)$ of Algorithm~\ref{alg:generic} to $(25, 15, 25, 0.25, 5, 0.01)$. 
Additionally, we set $B$ of Section~\ref{subsubsec: def of the energy loss} to $100$. For the filter function $f_\theta$, we employ a single layer neural network with rectifier activation function. We optimize the loss $L_\theta$ of line 7 in Algorithm~\ref{alg:generic} w.r.t. $\theta$ in $300$ epochs, where we use full-batch, and the learning rate of Adam optimizer is $0.001$. The computational environment is Apple M$1$ $16$ GB $8$ cores. The resulting Mapper graph with the labels by~\citet{davis2016modular} is shown in the right-hand side of Figure~\ref{fig: classical vs ours}. The running time is around 40 minutes.

For above experiments including the preprocessing, we choose hyperparameter values, which return the best result.

\begin{figure*}[!t]
      \begin{minipage}[!t]{.325\linewidth}
      \centering
        \includegraphics[width=0.9\hsize]{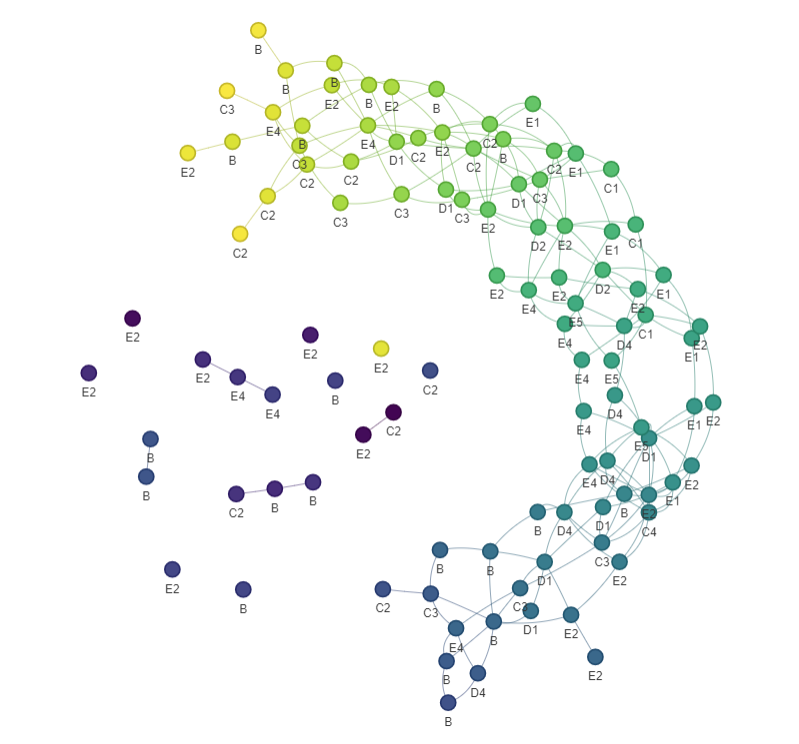}%
      \end{minipage} 
      \begin{minipage}[!t]{.3\linewidth}
      \centering
        \includegraphics[width=0.75\hsize]{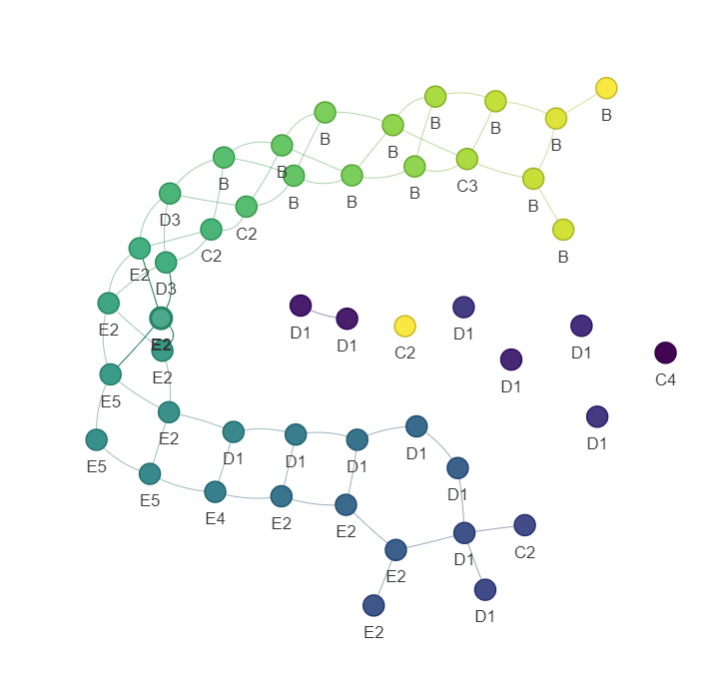}
      \end{minipage} 
      \begin{minipage}[!t]{.325\linewidth}
      \centering
        \includegraphics[width=0.9\hsize]{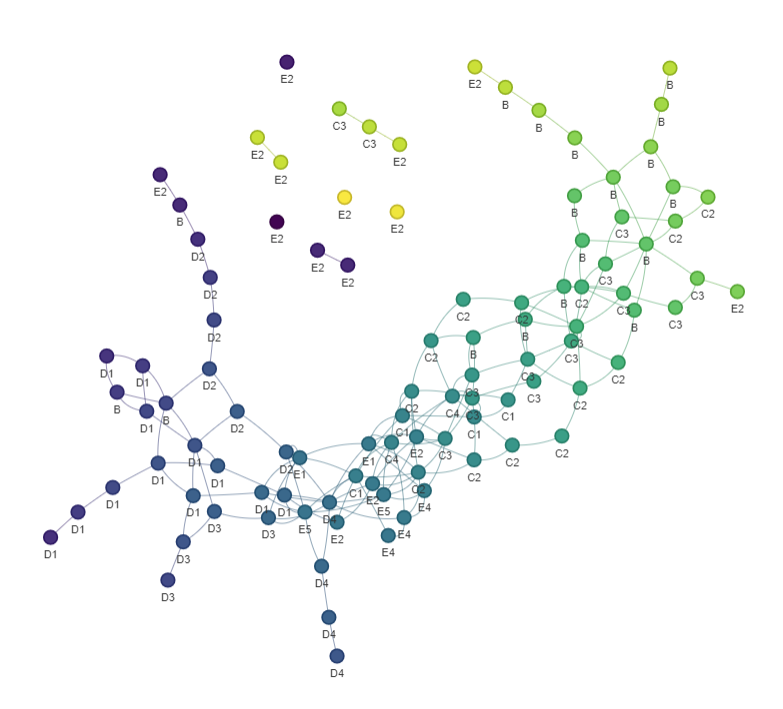}
      \end{minipage} 
\caption{
Three resulting Mapper graphs in ablation study. 
Left: The graph built on our clustering algorithm $\mathcal{A}_C$ in Section~\ref{subsubsec: def of clustering alg.} and the fixed filter function $f_\mathrm{mean}$. 
Middle: The graph build on k-means and our parameterized filter function $f_\theta$. 
Right: The graph by Algorithm~\ref{alg:generic} with $\lambda=0$. 
   }
\label{fig: other mapper results}   
\end{figure*}

\begin{figure*}[!t]
    \begin{minipage}[!t]{.245\linewidth}
      \centering
        \includegraphics[width=1\hsize]{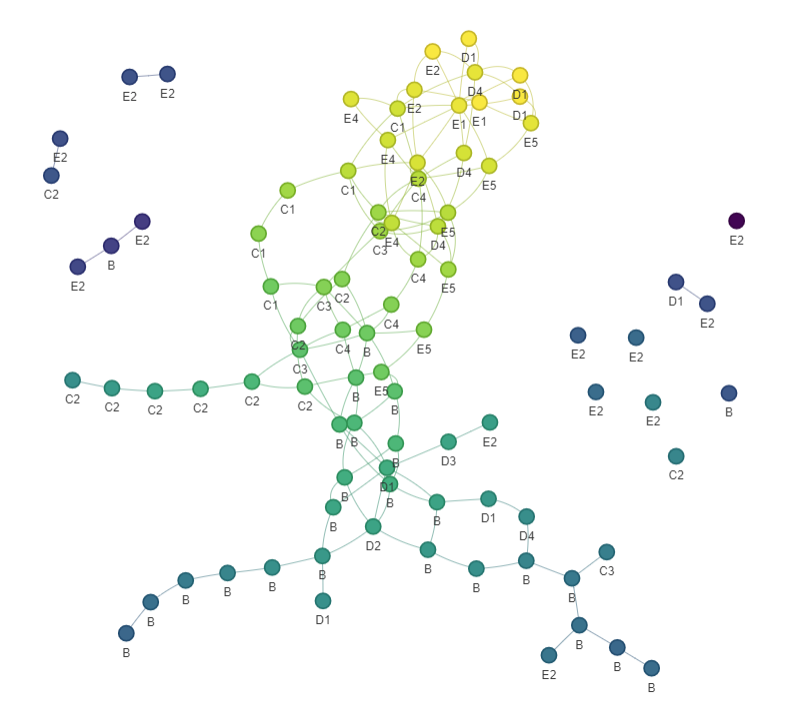}%
    \end{minipage} 
    \begin{minipage}[!t]{.245\linewidth}
      \centering
        \includegraphics[width=1\hsize]{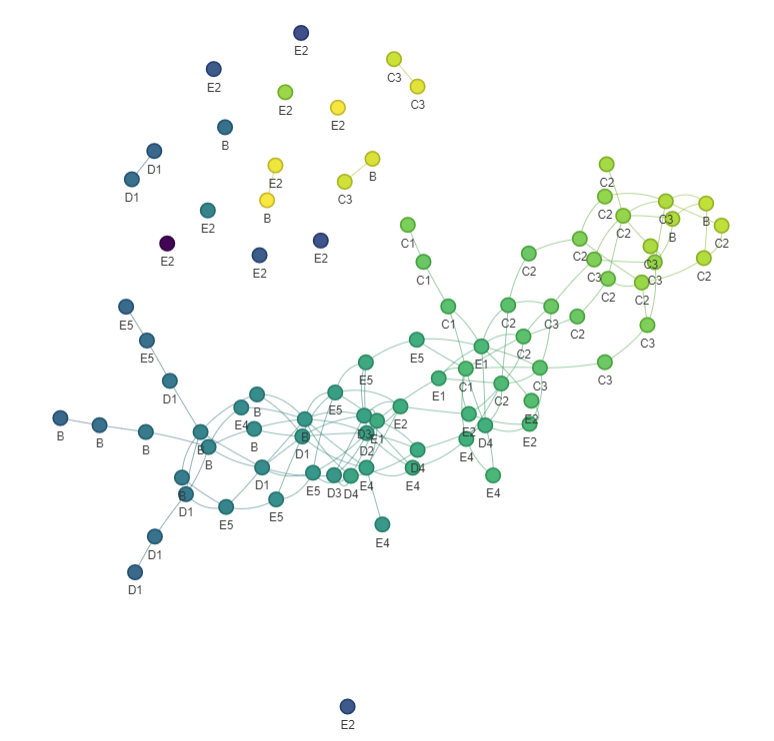}
    \end{minipage} 
    \begin{minipage}[!t]{.245\linewidth}
      \centering
        \includegraphics[width=1\hsize]{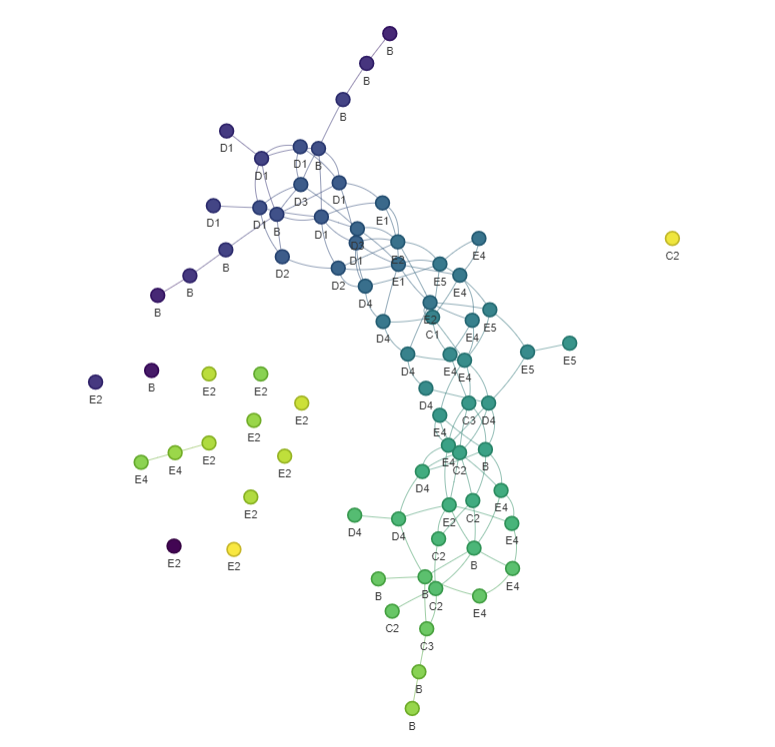}%
    \end{minipage} 
    \begin{minipage}[!t]{.245\linewidth}
      \centering
        \includegraphics[width=1\hsize]{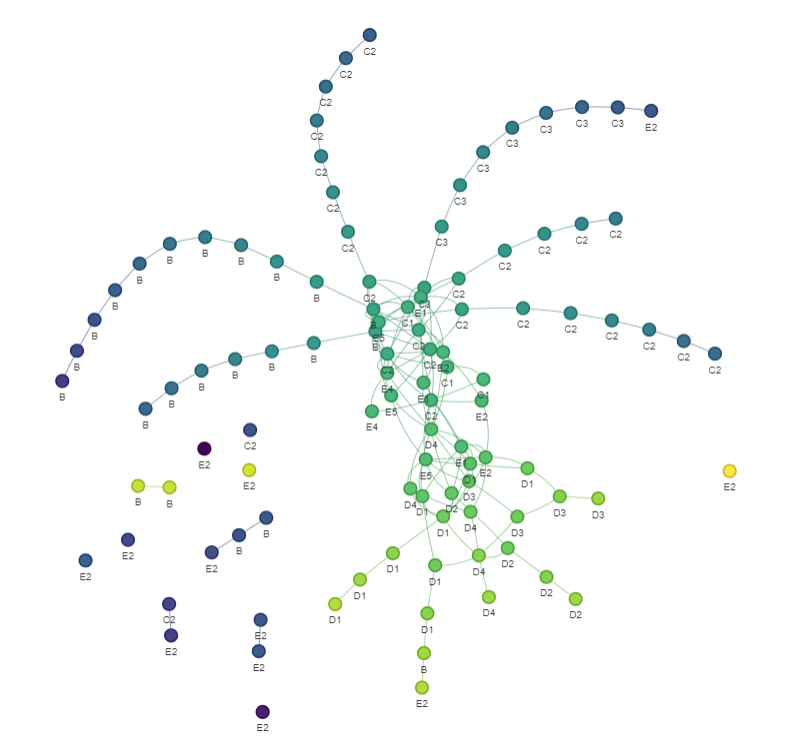}
    \end{minipage} 
\caption{
Four resulting Mapper graphs in robustness study. From right to left: The graphs constructed by Algorithm~\ref{alg:generic} with $(S,C)=(24,5)$, $(S,C)=(26,5)$, $(25, 4)$, and $(25,6)$, respectively.
}
\label{fig:resulting mapper graphs in robustness study}
\end{figure*}

\subsection{Analysis of Results}
\label{subsec: result and discussion}
The running time of our method (40 min) is sufficiently shorter than our main counterpart time of~\citet{yamazakiCryoTWIN} (240 min), although our computer (Apple M$1$ $16$ GB $8$ cores) is less powerful than~\citep{yamazakiCryoTWIN} (four NVIDIA V100 GPU
accelerators with two Intel Xeon Gold 6148 processors). However, in comparison with the classical Mapper algorithm, our method is much slower (40 min vs. 1 min), since ours optimizes the filter function.

As for the extraction of plausible pathways, from Figure~\ref{fig: classical vs ours}, we can observe that our Mapper graph contains all the four plausible 50S-ribosomal conformational pathways of~\citep[Figure~7]{davis2016modular}, which are the following $p_1$ to $p_4$:
\begin{description}
    \item[$p_1$:] 
    $\mathrm{B} \to \mathrm{D}1 \to \mathrm{D}2 \to 
    \mathrm{D}3 \to \mathrm{D}4 \to \mathrm{E}5$; see the red arrows,

    \item[$p_2$:]
    $\mathrm{B} \to \mathrm{D}1 \to \mathrm{E}1 \to \mathrm{E}2 \to \mathrm{E}4 \to \mathrm{E}5$,

    \item[$p_3$:]
    $\mathrm{B} \to \mathrm{C}2 \to \mathrm{E}1 \to \mathrm{E}2 \to \mathrm{E}4 \to \mathrm{E}5$; see the orange arrows,

    \item[$p_4$:]
     $\mathrm{B} \to \mathrm{C}2 \to \mathrm{C}3 \to \mathrm{C}1 \to 
    \mathrm{E}2 \to \mathrm{E}4 \to \mathrm{E}5$,
\end{description}
whereas the counterpart result by the classical method contains none of the four. 
In the pathway $p_1$, the structural label E3 is absent, because the label does not correspond to one of the top $25$-significant Gaussian components in the preprocessing.

Our Mapper graph moreover captures the important character of~\citep[Figure~7]{davis2016modular}: the labels C and D are well-separated overall in the graph. On the other hand, with two pairs ($\mathrm{D}2$, $\mathrm{E}1$) and ($\mathrm{D}4$, $\mathrm{E}4$), two labels in each pair are connected in our graph, while they should not according to~\citep[Figure~7]{davis2016modular}.

\begin{remark}
We are interested in a protein dataset that satisfies the following four conditions. First, the experts already have constructed the conformational pathway from the dataset, and the plausibility was well evaluated. Second, the pathway includes intermediate structures, whose occurrence probabilities are low. Third, the conformational diversity within the pathway is high. Fourth, the dataset is collected from cryo-EM; recall that we assume cryoTWIN for Algorithm~\ref{alg:generic} in this study. 
Note that analyzing such assembly pathways in a short running time is valuable, as ordinary methods (e.g., molecular dynamics) often struggle with the analysis.
To the best of our knowledge, one of the few datasets that satisfy the four conditions is the 50S-ribosomal dataset from EMPIAR-10076.
\end{remark}

\subsection{Ablation Study}
\label{subsec: ablation}
We evaluate our algorithm's performance in the three cases, using the latent variable set from the preprocessing. The first case is that we only replace the filter function $f_\theta$ in our algorithm with a fixed one. We consider the four fixed filter functions: $f_i$, $f_{\mathrm{min}}$, $f_{\mathrm{max}}$, and $f_{\mathrm{mean}}$. 
The filter $f_i$ is a projection from $\mathbb{R}^8$ to $\mathbb{R}$ which maps $z$ to $z_i$, where $z_i$ denotes the $i$-th entry of $z$. The filter functions $f_{\mathrm{min}}$, $f_{\mathrm{max}}$, and $f_{\mathrm{mean}}$ map $z \mapsto \min \{z_1, \dots, z_8\}, z \mapsto \max \{z_1, \dots, z_8\}$, and $z \mapsto \mathrm{mean} \{z_1, \dots, z_8\}$, respectively. The second case is that we only replace the clustering algorithm $\mathcal{A}_C$ in our algorithm with k-means. We set the number of clusters in k-means to $2$. The last case is that we remove the kNN regularizer from our algorithm, i.e., we set $\lambda$ in Algorithm~\ref{alg:generic} to $0$.

In the first case, among the eleven resulting Mapper graphs, we obtain Mapper graphs which contain three of the four plausible conformational pathways in~\citep{davis2016modular} at the best. One such the graph is shown on the left-hand side in Figure~\ref{fig: other mapper results}. For the second and third cases, none of the four conformational pathways is contained in the Mapper graph; see results of the second and third cases in the middle and right-hand side graphs of Figure~\ref{fig: other mapper results}, respectively. From those results, both the kNN regularizer $\mathrm{Reg}_{\theta}(G)$ and our designed clustering algorithm $\mathcal{A}_C$ in Algorithm~\ref{alg:generic} are important factors to extract the pathways from the latent distribution $P_\psi$, while the parameterized filter function $f_\theta$ also contributes to the extraction.

\subsection{Robustness Study}
\label{subsec: robustness}
We conduct a robustness study against the change of hyperparameters in Algorithm~\ref{alg:generic}. In this study, we focus on the following two important hyperparameters of the Mapper algorithm: the number of intervals $S$ and the maximum number of clusters $C$; see $S$ and $C$ in Algorithm~\ref{alg:generic}. Firstly, we change $S$ in $\{24, 25, 26\}$, while fixing the other hyperparameters in Algorithm~\ref{alg:generic} to the values described in the main experiment paragraph in Section~\ref{subsec: setting}. See the results with $S=24$ and $26$ at the first and second graphs from the left in Figure~\ref{fig:resulting mapper graphs in robustness study}, respectively. Secondly, we change $C$ in $\{4,5,6\}$, while fixing the other hyperparameters to the values used in the main experiment. See the results with $C=4$ and $6$ at the third and fourth graphs from the left in Figure~\ref{fig:resulting mapper graphs in robustness study}, respectively.

In the first graph of Figure~\ref{fig:resulting mapper graphs in robustness study}, one pathway $\mathrm{B} \to \mathrm{C}2 \to \mathrm{C}3 \to \mathrm{C}1 \to \mathrm{E}2 \to \mathrm{E}4 \to \mathrm{E}5$ is contained in the graph. In the second, two pathways $\mathrm{B} \to \mathrm{C}2 \to \mathrm{E}1 \to \mathrm{E}2 \to \mathrm{E}4 \to \mathrm{E}5$ and $\mathrm{B} \to \mathrm{C}2 \to \mathrm{C}3 \to \mathrm{C}1 \to \mathrm{E}2 \to \mathrm{E}4 \to \mathrm{E}5$ are contained in the graph. In the third, one pathway $\mathrm{B} \to \mathrm{D}1 \to \mathrm{E}1 \to \mathrm{E}2 \to \mathrm{E}4 \to \mathrm{E}5$ is contained in the graph. In the fourth, no complete pathway is contained in the graph, but two partial pathways $\mathrm{B} \to \mathrm{D}1 \to \mathrm{D}2 \to \mathrm{D}4 \to \mathrm{E}5$ and $\mathrm{B} \to \mathrm{C}2 \to \mathrm{C}3 \to \mathrm{C}1 \to \mathrm{E}2 \to  \mathrm{E}5$ are contained in the graph. From those analyses, our Mapper algorithm proves to be fairly sensitive to the change of hyperparameters.

\section{Conclusion and Future Work}
\label{sec:conclusion future work}
We propose a deep Mapper algorithm to extract plausible conformational pathways from the isometric latent distribution of cryoTWIN in short running time. In our numerical experiments, our method successfully extracts the well-recognized 50S-ribosomal pathways in shorter running time than the state-of-the-art method. 

One of our future work is to make our method to be more robust against change of the hyperparameters. Another future work is to apply our method to other datasets, and evaluate the results.

\appendix

\section{Complements for Mapper}
\label{sec: complements for mapper graph}
There are a number of choices to be made when computing a Mapper from a point cloud: 
\begin{description}
    \item[(i)]  the choice of intervals $(I_s)_{1\leq s\leq S}$: a popular practice when choosing a cover for the range of values of the filter is to choose intervals of the same length $\ell$, where no more than two intervals can intersect at once. The overlap rate $r$ between intervals is also constant (and is therefore less than $1/2$). By following the conditions above, we can tune the cover through $r$ and $\ell$. 

    \item[(ii)]  the choice of the clustering algorithm: any existing clustering algorithm is available to cluster the pre-image. One popular choice, however, is to construct a $\delta$-neighborhood graph and to look at its connected components as clusters. A $\delta$-neighborhood graph is a graph with vertices corresponding to each point in $\mathbb{X}_n$ and with an edge between two vertices if and only if the distance between the two points is less than $\delta$.
\end{description}

\begin{figure}[!t]
    \begin{minipage}[!t]{.45\linewidth}
      \centering
        \includegraphics[width=0.5\hsize]{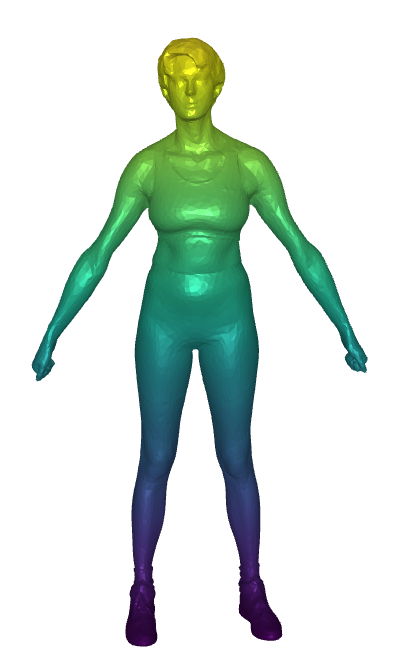}%
    \end{minipage} 
    \begin{minipage}[!t]{.45\linewidth}
      \centering
        \includegraphics[width=0.6\hsize]{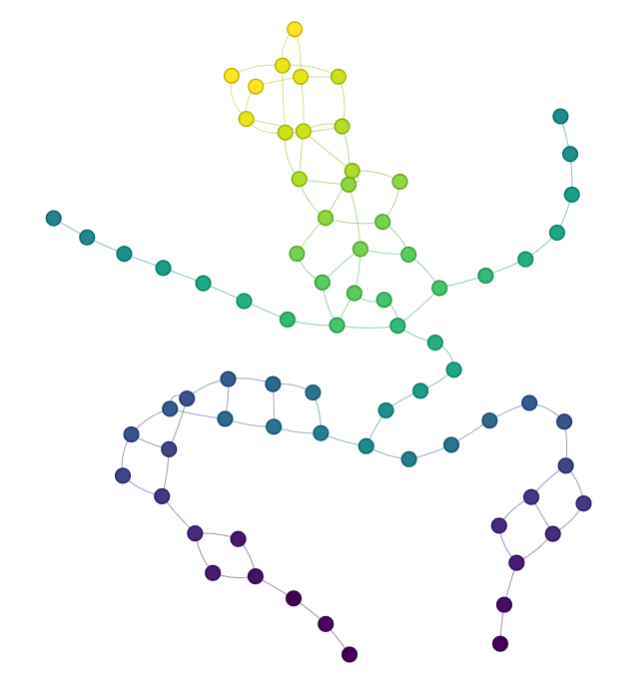}
    \end{minipage} 
\caption{
On the left: 3-dimensional mesh of a human shape colored in terms of height. On the right: Mapper graph computed using the vertices of the mesh, height as a filter, 25 equal intervals with 30\% overlap and k-means clustering with 3 clusters.}
\label{fig:mapper_example}
\end{figure}

A mesh of a human model in $3$-dimensions is in Figure \ref{fig:mapper_example}. We wish to compute the Mapper on the set of vertices of the mesh with the height function as our filter function. We choose to cover the range of height values in the manner we described above, with $25$ intervals of the same length with an overlap of $30\%$. For clustering, we use the k-means algorithm with $3$ clusters. The resulting graph is also represented in Figure \ref{fig:mapper_example}.
Here, we represent the mean value of the filter function on each cluster using the color of its vertex in the Mapper graph.
The Mapper graph we computed can be seen as a skeleton of the shape of a human : we can distinguish two legs, a torso and two arms.

\section{Proofs}
\label{sec: proofs}

\subsection{Local Distance Approximation}
\label{sec: details of lda}
For a smooth filter function $f\colon\mathbb{R}^d\rightarrow\mathbb{R}$, Taylor's theorem in the first order can be written for each coordinate $i\in\{1,\dots ,d\}$ as
$\forall z\in \mathbb{R}^d,$ and $h\in\mathbb{R}$:
$$f(\Tilde{z})=f(z)+(\nabla f(z)_i)\cdot h+o(|h|),$$ 
where $\Tilde{z}=(z_1,\dots ,z_{i-1},z_{i}+h, z_{i+1},\dots ,z_d)$ and $\nabla f(z)_i$ is the $i-$th coordinate of the gradient of $f$ in $z$.
As such, we can approximate local distances by using the gradient of the filter function, see Algorithm \ref{alg:locaprox}. Note that the approximation can be pushed further by using Taylor's theorem to higher orders, however this can be more costly to compute. 
\begin{algorithm}[!t]
\caption{Local Distance Approximation (LDA)}
\label{alg:locaprox}
\KwInput{

        $z,\,z'\in\mathbb{R}^d$: points,

        $f$: a filter function,

        $\nabla f(z)$: the gradient of $f$ at $z$.
        }
\For{$i\in \{1, \dots, d\}$}
    {$\Tilde{z} \gets(z_1, \dots, z'_i, \dots, z_d)$, and then 
    $\partial f_i = f(\Tilde{z}) - f(z)$.
    }
\Return $\left\lVert\left(\frac{\partial f_1}{\nabla f(z)_1},\dots,\frac{\partial f_i}{\nabla f(z)_i},\dots,\frac{\partial f_d}{\nabla f(z)_d}\right)\right\rVert_2$.
\end{algorithm}

\subsection{Proof of Equation~\eqref{eq1}}
\label{append: proof eq1}
\begin{proof}
We firstly show that
$\{S_{l}\}_{l\geq2}$ 
is lower bounded by $0$ and non-increasing.
Let 
$\{z_i\}_{i=1}^{l-1}\subseteq\mathbb{R}^d$, 
and consider 
$\{z'_i\}_{i=1}^{l}\subseteq\mathbb{R}^d$ 
such that 
$z'_1=z_1$ 
and for any 
$i \in \{2,\dots,l\}$,  $z'_i=z_{i-1}$ holds. 
Note that we always have $z_{l}=z'_{l+1}=z_\infty$. 
We then have
\begin{equation*}
    \sum_{i=0}^{l-1}\frac{1}{P\left(\frac{z_i+z_{i+1}}{2}\right)}\left\Vert z_{i+1}-z_i \right\Vert_2
    =\sum_{i=0}^{l}\frac{1}{P\left(\frac{z'_i+z'_{i+1}}{2}\right)}\left\Vert z'_{i+1}-z'_i \right\Vert_2.
\end{equation*}
Accordingly, we have 
\begin{equation*}
\left\{\sum_{i=0}^{l-1}\frac{1}{P\left(\frac{z_i+z_{i+1}}{2}\right)}\left\Vert z_{i+1}-z_i \right\Vert_2,\,\{z_i\}_{i=1}^{l-1} \subseteq \mathbb{R}^d \right\}  
\subseteq \left\{\sum_{i=0}^{l}\frac{1}{P\left(\frac{z_i+z_{i+1}}{2}\right)}\left\Vert z_{i+1}-z_i \right\Vert_2,\,\{z_i\}_{i=1}^{l}\subseteq\mathbb{R}^d \right\}.
\end{equation*}
Therefore 
$S_{l+1}\leq S_{l}$ 
holds by the definition of the infimum.
We showed that 
$\{S_{l}\}_{l \geq2}$ 
is convergent, and we furthermore have

\begin{equation*}
\lim_{K\to\infty}S_K=\inf_{K\geq 2}\{S_K\}.
\end{equation*}

We now show that this limit is exactly $S^\ast$.
We firstly have 
$S^\ast\leq\lim_{K\to\infty}S_{l}$ 
because for any 
$l \geq 2$, $S^\ast\leq S_{l}$ 
holds.
Let 
$l \in \mathbb{N}$ such that $l>2$,
and fix a discrete path 
$\{z_i\}_{i=1}^{l-1}\subseteq\mathbb{R}^d$. 
Then let us consider the piecewise constant function 
$g \colon [0,1] \to \mathbb{R}^d$ 
such that for each 
$i \in \{0,\dots,l-1\}$
and for any 
$t\in[\frac{i}{l},\frac{i+1}{l})$, 
$g(t)=z_i$ and $g(1)=z_\infty$
hold.
We then have
\begin{equation*}
I_g=\sum_{i=0}^{l-1}\frac{1}{P\left(\frac{z_i+z_{i+1}}{2}\right)}\left\Vert z_{i+1}-z_i \right\Vert_2.
\end{equation*}
This is because the sequence of Riemann sums, which limit is by definition equal to $I_g$, is stationary after rank $l$ and its terms are equal to the sum above.
Since $g\in \mathrm{PC}^0\left([0,1],\mathbb{R}^d\right)$,
\begin{equation*}
S^\ast\leq\sum_{i=0}^{l-1}\frac{1}{P\left(\frac{z_i+z_{i+1}}{2}\right)}\left\Vert z_{i+1}-z_i \right\Vert_2
\end{equation*}
holds for all 
$\{z_i\}_{i=1}^{l-1}\subseteq\mathbb{R}^d$.
Hence we conclude that 
$S^\ast\leq S_{l}$. 

We secondly show $\lim_{l\to\infty}S_{l}\leq S^\ast$.
Let 
$f\in \mathrm{PC}^0\left([0,1],\mathbb{R}^d\right)$.
By definition of $I_f$, 
\begin{equation*}
\forall \epsilon>0, \exists n_{f,\epsilon}\in\mathbb{N} \text{ such that } \forall n\geq n_{f,\epsilon}, 
\end{equation*}
\begin{equation*}
\left|\sum_{i=0}^{n-1}\frac{1}{P\left(\frac{f(\frac{i}{n})+f(\frac{i+1}{n})}{2}\right)}\left\Vert f\left(\frac{i+1}{n}\right)-f\left(\frac{i}{n}\right) \right\Vert_2 - I_f\right|\leq\epsilon.
\end{equation*}
Hence, 
\begin{equation*}
\forall \epsilon>0, \exists n_{f,\epsilon}\in\mathbb{N} \text{ such that }, 
S_{n_{f,\epsilon}}\leq I_f+\epsilon.
\end{equation*}
Moreover, since $\lim_{l\to\infty}S_{l} =\inf_{l \geq 2}\{S_{l}\}$, 
$\lim_{l\to\infty}S_{l}\leq I_f+\epsilon$ follows for any $\epsilon > 0$.
Therefore, for any $f\in \mathrm{PC}^0\left([0,1],\mathbb{R}^d\right)$,
$\lim_{l\to\infty}S_{l}\leq I_f.$
holds.
Therefore, we conclude that $\lim_{l \to\infty}S_{l}\leq S^\ast$. Finally we have
\begin{equation*}
S_{l}\xrightarrow[l \to \infty]{}S^\ast.
\end{equation*}
\end{proof}

\subsection{Proof of Equation~\eqref{eq2}}
\label{append: proof eq2}
\begin{proof}
For the rest of the proof, we fix $l\in\mathbb{N}$ such that $l>2$.\\
We firstly work at a fixed elementary event in the probability space where our random sequences are defined. Note that 
$\{S_{l}^{(N)}\}_{N\geq1}$
is lower bounded by $0$ and non-increasing.
Let $N\geq l$. 
Since $\{z_j\}_{j=1}^N\subseteq\{z_j\}_{j=1}^{N+1}$, we have that $S_{l}^{(N+1)}\leq S_{l}^{(N)}$. 
Therefore, $\{S_{l}^{(N)}\}_{N\geq1}$ converges and its limit is
\begin{equation*}
\lim_{N\to\infty}S_{l}^{(N)}=\inf_{N\geq 2}\{S_{l}^{(N)}\}.
\end{equation*}
We now have that 
$\{z_j\}_{j=1}^N\subseteq\mathbb{R}^d$ 
for every $N\geq1$. 
This shows that for every realization of the random sequence
\begin{equation*}
\forall N\geq1\,:\, S_{l}\leq S_{l}^{(N)}
\end{equation*}
and 
\begin{equation*}
S_{l}\leq \lim_{N\to\infty}S_{l}^{(N)}.
\end{equation*}
Let us consider the following function $S$ defined as
\begin{align*}
S\colon {(\mathbb{R}^d)}^{l-1} & \longrightarrow\mathbb{R},\\
\{z_i\}_{i=1}^{l-1}&\longmapsto \sum_{i=0}^{l-1}\frac{1}{P\left(\frac{z_i+z_{i+1}}{2}\right)}\left\Vert z_{i+1}-z_i \right\Vert_2.
\end{align*} 
Let $k\in\mathbb{N}$. We fix a discrete path 
$z=\{z_i\}_{i=1}^{l-1} \subseteq \mathbb{R}^d$, such that $S(z)\leq S_l+\frac{1}{2k}$.
As stated in Theorem \ref{thm}, for any 
$i\in\{0,\dots,l-1\}$,  $z_{i+1}\neq z_i$ and $P$ is smooth. 
This means that $S$ is also smooth in some open ball $B(z,\epsilon_z)$ centered at $z$ with radius $\epsilon_z$.
Since $S$ is smooth in $B(z,\epsilon_z)$, it is locally Lipschitz continuous in the closed ball $\Bar{B}(z,\frac{\epsilon_z}{2})$ for some constant $L_z$. 
This comes from the mean value theorem and the fact that the gradient of $S$ is bounded on the compact set 
$\Bar{B}(z,\frac{\epsilon_z}{2})$.\\
Then for any $z'\in \Bar{B}(z,\frac{\epsilon_z}{2})\cap \Bar{B}(z,\frac{1}{2k\cdot L_z})$ 
we have
\begin{align*}
    \mid S(z)-S(z')\mid&\leq L_z\Vert z-z'\Vert_2 \\
    &\leq\frac{1}{2k}.
\end{align*}
Furthermore, $P$ is strictly positive almost everywhere. 
This means that with the notation $A_z=\Bar{B}(z,\frac{\epsilon_z}{2})\cap \Bar{B}(z,\frac{\epsilon}{L_z})$, 
\begin{equation*}
\int_{A_z} P(v_1) \cdots P(v_{l-1})\mathrm{d}v_1 \cdots \mathrm{d}v_{l-1}>0
\end{equation*}
holds.
Therefore, almost surely, there exists a rank $N_z$ after which there exists $z'\in A_z$ such that $z'\subseteq\{z_j\}_{j=1}^{N_z}$.
Then the following proposition holds:
\begin{equation*}
\forall k\in\mathbb{N}, \exists N_z\in\mathbb{N} \text{ almost surely such that } S_{l}^{(N_z)}\leq S_l+\frac{1}{k},
\end{equation*}
which implies that for any $k\in\mathbb{N}$, $|\lim_{N\to\infty}S_{l}^{(N)}-S_l|\leq \frac{1}{k}$ in a subset of measure $1$ of the probability space.\\
The countable intersection of almost sure events being also almost sure, we have
\begin{equation*}
S_{l}^{(N)}\xrightarrow[N \to \infty]{a.s}S_{l}.
\end{equation*}
\end{proof}

\section{Isometric Property of cryoTWIN}
We briefly review the property, since it bridges a filtered space of deep mapper and a space of protein conformation as 3D electronic density map (i.e., an output space of cryoTWIN), via the latent space.

After training cryoTWIN by Equation~\eqref{eq: cryotwin objective}, the latent space can be isometric to the conformational space in the following sense:
\begin{equation}
\label{eq: isometric property}
    \forall (z, \delta_1, \delta_2);\;\left\langle ( z + \delta_1 ) - z,  ( z + \delta_2 ) - z \right\rangle 
    \propto
    \left\langle \hat{V}_{z + \delta_1} - \hat{V}_{z},  \hat{V}_{z + \delta_2} - \hat{V}_{z} \right\rangle,
\end{equation}
where $z$ is the latent variable, $\delta$ is an infinitesimal vector, and $\hat{V}_{z}$ is the reconstructed 3D density map. Once Equation~\eqref{eq: isometric property} holds, we can immediately obtain two equations as follows:
\begin{equation*}
\|z - z'\|_2 \approx 0 \Rightarrow \|z - z'\|_2 \propto \left\| \hat{V}_z - \hat{V}_{z'} \right\|_2,\;\mathrm{and}\; 
P_{\psi} (z) \propto  p\left(\hat{V}_z\right),
\end{equation*}
where $p$ is the conformational distribution.

\section{FSC Evaluations in Preprocessing}
\label{subsec: FSC Evaluations in Preprocessing}
\begin{figure*}[!t]
    \begin{minipage}[!t]{.225\linewidth}
      \centering
        \includegraphics[width=1\hsize]{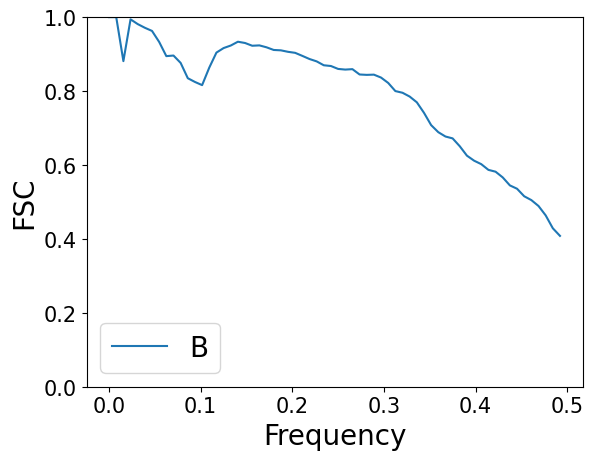}%
    \end{minipage} 
    \begin{minipage}[!t]{.225\linewidth}
      \centering
        \includegraphics[width=1\hsize]{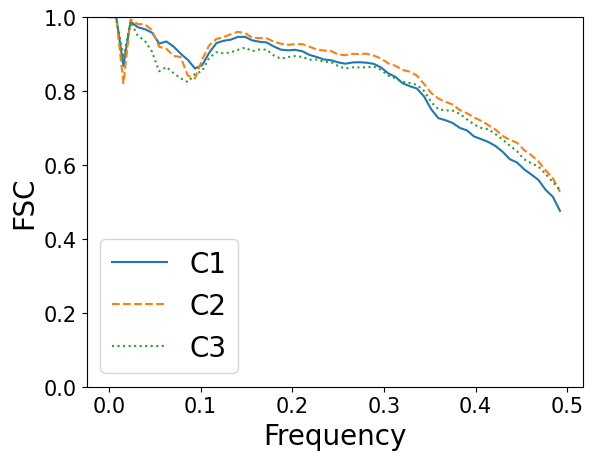}
    \end{minipage} 
    \begin{minipage}[!t]{.225\linewidth}
      \centering
        \includegraphics[width=1\hsize]{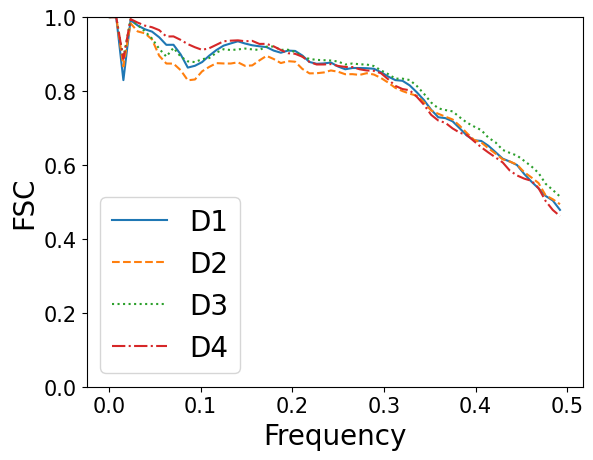}%
    \end{minipage} 
    \begin{minipage}[!t]{.225\linewidth}
      \centering
        \includegraphics[width=1\hsize]{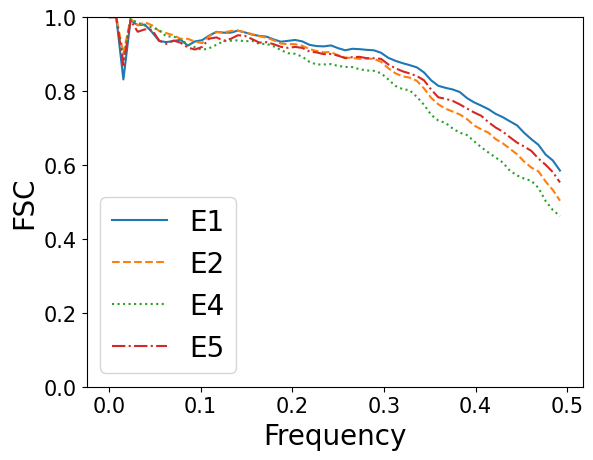}
    \end{minipage} 
\caption{
Figures of FSC corresponding to labels B, C, D, and E from left to right respectively.
}
\label{fig:resulting fsc graphs in preprocessing}
\end{figure*}

In order to evaluate our 3D density maps by FSC, we employ a kind of pseudo true 3D density maps, which are publicly available in the URL\footnote{\url{https://zenodo.org/records/4355284\#.YCq_dI9Kj0o}}. 
The pseudo true density maps are computed by a machine learning technique named cryoDRGN of~\citep{Zhong2020Reconstructing}. For the FSC evaluations, we down-sample the size of the 3D density maps in the URL from $256\times 256\times 256$ to $128\times 128\times 128$.  The results are shown in Figure~\ref{fig:resulting fsc graphs in preprocessing}.

\bibliographystyle{apalike}
\bibliography{refs-camera-ready}

\end{document}